# On-chip generation, routing and detection of quantum light

G. Reithmaier[1], M. Kaniber*[1], F. Flassig[1], S. Lichtmannecker[1], K. Müller[1,3], A. Andrejew[1], J. Vučković[3], R. Gross[2,4] and J. J. Finley*[1,4]

[1]Walter Schottky Institut und Physik Department, Technische Universität München, Am Coulombwall 4, 85748 Garching, Germany.

[2]Walther-Meißner-Institut, Bayerische Akademie der Wissenschaften und Physik-Department, Technische Universität München, 85748 Garching, Germany.

[3]E. L. Ginzton Laboratory, Stanford University, Stanford, CA 94305, USA.

[4]Nanosystems Initiative Munich (NIM), Schellingstraße 4, 80799 München, Germany.

*Correspondence to finley@wsi.tum.de or kaniber@wsi.tum.de

**Semiconductor based photonic information technologies are rapidly being pushed to the quantum limit where non-classical states of light can be generated, manipulated and exploited in prototypical quantum optical circuits.[1,2] Here, we report the *on-chip* generation of quantum light from individual, resonantly excited self-assembled InGaAs quantum dots, efficient routing over length scales ≥ 1 mm via GaAs ridge waveguides and in-situ detection using evanescently coupled integrated NbN superconducting single photon detectors fabricated on the same chip. By temporally filtering the time-resolved luminescence signal stemming from single, resonantly excited quantum dots we use the prototypical quantum optical circuit to perform time-resolved excitation spectroscopy on single dots and demonstrate resonant fluorescence with a line-width of 10 ± 1 μeV; key elements needed for the use of single photons in prototypical quantum photonic circuits.**

Non-classical light can be generated by resonant excitation of individual quantum emitters to scatter single photons with coherence properties determined primarily by the excitation laser source[3]. By carefully tailoring the local optical mode density experienced by the emitter, coherent and incoherent non-classical light can be preferentially routed into low loss waveguide modes and distributed on a chip into a quantum photonic circuit[4,5,6,7]. In recent years strong quantum light-matter couplings in such photonic nanostructures have been shown to produce *effective* interactions between photons leading to remarkable phenomena such as photon blockade[8,9,10], needed for optical transistors[11,12], photonic quantum gates[13,14], and ultrafast optical switching with only a few control photons[15]. Whilst the generation and routing of quantum light on a semiconductor chip has already been demonstrated by several groups[16,17,18], the ability to generate and route single photons on-chip and detect them *in-situ* with near unity quantum efficiency[19] would represent a major step towards the realization of semiconductor based quantum optical circuits. Hereby, superconducting single photon detectors (SSPDs) are particularly useful due to their very high detection efficiencies[20,21,22,23], low dark count rates[24], sensitivity from the visible to the IR[25] and picosecond timing resolution[26,27]. Single photon detection in such devices arises from the formation of a normal conducting hotspot in the thin superconducting nanowire[28]. Since the bias current is slightly sub-critical (typically $\sim 0.95 \times I_{crit}$), the local heating arising from single photon absorption results in the breakup of Cooper pairs, locally switching of the nanowire to a normal conducting state thus giving rise to a measurable voltage pulse in the external readout circuit. The possibility to integrate SSPDs onto dielectric[1] and plasmonic[2,29,30] waveguides facilitates evanescent coupling to waveguide photons, pushing the single-photon detection efficiency towards unity. Both the generation of cluster states of photonic qubits for one-way quantum computation[31] and the measurement based teleportation schemes[32] rely on having such near perfect detection efficiency.

As depicted schematically in fig 1a the samples investigated consist of a linear GaAs (core) – AlGaAs/air (cladding) multimodal ridge waveguide into which a single layer of optically active self-assembled InGaAs quantum dots (QDs) is embedded at its midpoint, $125 \pm 5$ nm below the surface. A single-frequency tuneable diode laser (line-width $<$ 1 MHz, modulated at $\sim 3$ GHz to produce $290 \pm 70$ ps pulses at a 3 % duty cycle) is focused onto the waveguide surface producing a diffraction limited spot with $\sim 3 - 5$ self-assembled QDs within the focal volume (fig 1a). The light generated and scattered by the excited QDs is then guided along the ridge waveguide over a distance $\Delta x = 1040 \pm 1$ μm and evanescently coupled into the nanowires of an integrated SSPD as shown in the scanning electron microscopy image in fig 1a. By varying $\Delta x$ whilst recording the QD photoluminescence (PL) using the integrated SSPD[33] we determined the propagation losses to be $6.6 \pm 0.5$ dB/mm. A simulated photonic mode profile for a $w_{WG} = 20$ μm wide waveguide is shown as an inset of fig 1a, demonstrating the strong confinement of light in the waveguide; this gives rise to near-unity absorption of waveguide-photons in the active detection-region[34]. This sample geometry allows the detection of confocal QD luminescence emitted via the top waveguide surface and the simultaneous investigation of the temporally resolved QD response guided along the waveguide using the integrated SSPD. Therefore, this prototypical quantum optical circuit provides a test bed for all-optical generation, distribution and detection of non-classical light on the photonic chip.

Firstly, we performed off-chip detected confocal PL measurements for various positions along the waveguide in order to identify regions with only a few sharp emission lines originating from individual dots. Typical results of such *off-chip* detected confocal PL measurements are presented in fig 1(b) that compares emission spectra recorded from one such position with the laser frequency tuned into the wetting layer at $E_{WL} = 1347.7$ meV (blue curve) and quasi-resonant excitation via an excited QD resonance, labelled $X_2^{**}$ in fig 1(b), at $E_{X_2^{**}} =$ 1302.8 meV (red curve). For this excitation position and both excitation energies we observe pronounced emission from the same line labelled $X_1$ in fig 1(b) at $E_{X_1} = 1263.5$ meV and weaker emission from the line labelled $X_2$ at $E_{X_2} = 1268.2$ meV. As shown in the supplementary material (section S2), the intensity of both emission lines increases linearly with the excitation power ($P$) according to $I \propto P^m$ with exponents of $m = 1.00 \pm 0.03$ and $1.02 \pm 0.05$, respectively, identifying both features as arising from single exciton transitions[35]. We will continue to probe the absorption properties of $X_1$ and $X_2$ using PL – excitation (PLE) spectroscopy later in figs 2(a) and 2(b) with both *on-chip* detection via the integrated SSPD and *off-chip* detection via an imaging monochromator (see methods). The bottom axis of fig 1(b) shows the energy difference $\Delta E_{det} \equiv E_{det} - E_{X_1}$ between the detection energy ($E_{det}$) and the $X_1$ ground state emission energy. Whereas the emission lines $X_1$ and $X_2$ are attributed to ground state transitions of two different QDs, the lines $X^*_{1/2}$, $X^{**}_{1/2}$ and $X^{***}_1$ marked in fig 1(b) are identified as arising from excited QD states, an attribution supported by the PLE measurements discussed below.

Performing *on-chip* PLE spectroscopy using the integrated SSPD is complicated by the fact that it is sensitive to *all* light propagating within the waveguide mode (i.e. no spectral filtering). Moreover the SSPD detects any background signal originating from laser stray light scattered at the back surface of the sample as well as laser light that is directly coupled into the waveguide. In an effort to reduce the impact of scattered laser light we polished the back surface of the sample and sputtered an amorphous silicon absorber layer to reduce the index contrast and, thereby, the impact of scattering. Furthermore, this layer also absorbs the scattered excitation laser light (see methods). In addition, we utilised pulsed laser excitation in combination with temporal filtering of the SSPD signal to separate the directly scattered laser light from the QD luminescence. Fig 1(c) shows a typical example of the time-resolved SSPD signal following excitation next to the waveguide where all QDs have been completely removed by etching. In this case, only scattered laser light reaches the SSPD and the time-resolved SSPD signal reveals a Gaussian instrument response function (IRF). The measured IRF FWHM of 290 ± 70 ps reflects the temporal resolution of the high-frequency detection electronics, since the SSPD response is much faster (~ 70 ps)[34]. In strong contrast, upon exciting the waveguide region containing QDs via the underlying wetting layer we observe an exponentially decaying SSPD signal that can clearly be seen in fig 1(d). The deconvolution of the observed SSPD signal with the IRF reveals a clear mono-exponential decay with a lifetime of $\tau_{QDs} = 0.80 \pm 0.07$ ns, typical for the InGaAs QDs in this sample[35]. In order to completely separate the QD signal from the laser stray light, the detected time-resolved SSPD signal was temporally integrated for times $\geq 0.5$ ns onwards as illustrated by the yellow shaded region in fig 1(d).

This technique to distinguish the on-chip QD-PL signal from scattered laser light reaching the SSPD facilitates the investigation of the PLE response of the $X_1$ – transition. Thereby, the excess laser energy $\Delta E \equiv E_{exc} - E_{X_1}$ was tuned over a range of 17.0 meV $\leq \Delta E \leq$ 52.5 meV in steps of 0.1 meV. The corresponding time-resolved transients for on-chip detection and the $\Delta E$ dependent *off-chip* multi-channel PLE data showing the $X_1$ and $X_2$ peaks are presented in figs 2(a) and 2(b), respectively. For large excess excitation energies $\Delta E_{exc} = 47.5 \rightarrow$ 52.5 meV a strong on-chip signal is observed with closely spaced sharp lines in PLE which we attribute to WL 0D-2D bound-continuum transitions[36] (see figs 2(a) and 2(b)). Moving to more resonant excitation energies ($\Delta E_{exc} = 46.0 \pm 1.0$ meV), we simultaneously observe a suppression of luminescence in both on- and off-chip measurements. This observation reflects the absence of continuum transitions[36] for more resonant excitation. For excess laser energies in the range $\Delta E_{exc} \leq 44.0$ meV the PLE spectrum exhibits several discrete absorption resonances on a weak background (see fig 2(a) and 2(b)). The three resonances labelled $X_1^{**}$, $X_2^{**}$ and $X_1^{***}$ in fig 2(a) will be focussed on later in the discussion related to fig 2(c). Most interestingly, we observe a group of long time-transients labelled $X^*_{1/2}$ for $\Delta E_{exc}<$ 25.0 meV in fig 2(a) and corresponding weak QD emission features as shown in fig 2(b). These states are tentatively attributed to p-shell transitions of the QD emission lines $X_1$ and $X_2$, consistent with the expected orbital quantization energy $\Delta E_{s-p} \sim 20$ meV for such InGaAs QDs[37]. This identification is further supported by the fact that no additional PLE resonances are observed in the on-chip data for $\Delta E_{exc} < 17.5$ meV down to the s-shell transition. We note that the increasing background observed in the off-chip measurement for $\Delta E_{exc} < 17.5$ meV (figs 2(b) and 2(c)-bottom panels) arises from scattered laser light at the sample.

We continue by focusing on the energy range between $20.0$ meV $< \Delta E_{exc} < 45.0$ meV, where several pronounced and clearly separated excited state resonances for both on- and off-chip detection geometries are observed. The lower two panels of fig 2(c) show the off-chip PLE spectra detecting at the energies of $X_1$ and $X_2$, respectively, whereas the upper panel shows the time integrated on-chip SSPD signal for $t \geq 0.5$ ns to exclude the scattered laser light. When comparing the on- and off-chip PLE measurements presented in fig 2(c), several common features can be identified as highlighted by the grey shaded regions. For example, at $\Delta E_{exc} = 20.0$ meV a clear resonance is observed in the on-chip signal shown in the upper panel of fig 2(c), whilst simultaneously a clear resonance is observed in the PLE spectrum detecting on $X_2$ (fig 2(c) – middle panel). In total we focus on two distinct PLE resonances ($X^*_2, X^{**}_2$) for $X_2$ and on the three resonances $X^*_1, X^{**}_1$ and $X^{***}_1$ for $X_1$ and marked them in fig 2(c). A high-resolution scan with on-chip detection of $X^*_1$ is presented in fig 2(d) showing a Lorentzian shaped resonance at $\Delta E_{exc} = 22.24$ meV. In addition to the highlighted resonances, a number of weaker features were detected in the confocal PL data between $23.0$ meV $< \Delta E_{exc} < 30.0$ meV and found to be correlated with resonances observed in the on-chip PLE in the same energy interval. In summary, common QD resonances are observed both in off- and on-chip PLE measurements, indicating that the integrated SSPD has sufficient sensitivity to discriminate emission from an individual quantum emitter.

After demonstrating that our on-chip integrated SSPDs are capable of measuring on-chip PLE, whilst simultaneously collecting off-chip PLE of QD excited state resonances, we continue to demonstrate *resonant fluorescence* (RF) from the s-shell transition $X_1$. Hereby, we fine-tuned the excitation laser in steps of $\delta E = 1$ µeV across $X_1$ and simultaneously recorded on-chip time-resolved transients for each laser excitation energy. As discussed in the supplementary material (section S3), resonant fluorescence in such patterned photonic systems typically requires use of a second, weak non-resonant laser tuned above the bandgap of the GaAs waveguide ($E = 1.534$ eV, $P = 0.06 \pm 0.02$ W/cm$^2$) to continuously create free carriers in the vicinity of the QD under study and, thereby, stabilize the fluctuating electrostatic environment[38,39]. In fig 3(a), we present time-resolved transients recorded as a function of detuning $\Delta E_{exc} = E_{exc} - E_{X_1}$ whilst scanning the laser across the s-shell resonance $X_1$. We clearly observe resonantly excited QD time transients over an energy range $\Delta E_{exc} = \pm 20$ µeV with respect to the QD exciton ground state energy $E_{X_1}$. The observed signal exponentially decays with a lifetime of $0.70 \pm 0.07$ ns clearly showing that it arises from QD emission. The on-chip recorded RF signal, integrated over a temporal range of $0.5 - 5$ ns, is presented in fig 3(b) as a function of excitation energy $\Delta E_{exc} = \pm 60$ µeV exhibiting a pronounced peak at $\Delta E_{exc} = 0$ µeV. Fitting the data with a Lorentzian line shape ( $2A/\pi \times \Delta_{FWHM}/(4 \Delta E_{exc}^2 + \Delta_{FWHM}^2)$, with the peak area *A*) we extracted a line width of $\Delta_{FWHM} = 10.2 \pm 1.3$ µeV, in excellent agreement with values recently reported for *off-chip* detected RF from QDs embedded in a free standing single-mode ridge waveguide[39]. The inset of fig 3(b) shows the linear power dependence of the temporally integrated RF signal detected at $\Delta E_{exc} = 0$ µeV as a function of the excitation laser power. The observed linear dependence is expected for the $290 \pm 70$ ps pulses used, since the excitation pulse duration exceeds the QD coherence time of $T_2 = \hbar/\Delta E = 65 \pm 8$ ps and thereby coherent Rabi dynamics are expected to be weak. Similar resonant fluorescence measurements were made on several different dots within the waveguide with similar results to those presented in figs 3(a) and 3(b). Fig 3(c) shows one example RF spectrum from another dot labelled $X_3$ emitting at 1296.6 meV (PL spectra presented in the supplementary material section S2). Using a Lorentzian fit (red solid line) we extract the line width of $X_3$ to be $13.0 \pm 4.3$ µeV, identical to the RF signal obtained from $X_1$ within experimental accuracy. In order to demonstrate the quantum nature of the on-chip detected light we performed off-chip measurements of the photon statistics of $X_3$. Since emission is predominantly into the WG mode, the signal recorded in off-chip detection is weak. However, this particular QD emission line exhibits a very clean ground state emission signal $S$ with little background $B$, resulting in an emission coefficient[40] of $\delta = S/(S + B) = 0.92 \pm 0.05$. When filtering $X_3$ using a monochromator and sending it into a Hanbury-Brown and Twiss setup, we measured the autocorrelation function $g^{(2)}(\tau)$, as presented in fig 3(d). Here, the background corrected[40] coincidence counts detected from the QD emission are plotted as a function of time delay. The red solid line shows a deconvoluted fit with $g^{(2)}(0) = 0.36 \pm 0.07$ proving quantum nature of the emission from this dot. The observation of antibunched emission from $X_3$, that exhibits a clear RF signature in our on-chip measurements, provides strong evidence that the on-chip RF signal observed in our experiments represents non-classical light in the prototype quantum optical circuit shown in fig 1(a).

In summary, we presented the creation, routing and detection of quantum light on a single chip. By temporally filtering the on-chip detected, time-resolved luminescence signal, we suppress the background generated by the excitation laser, demonstrating that specific QD excited state resonances can be observed in both on- and off-chip detection geometries. Applying this time-filtering technique, we demonstrated resonant fluorescence from a single

QD with a line-width of 10.2 ± 1.3 µeV, guided in the optical modes of a GaAs ridge waveguide and efficiently detected via evanescent coupling to an integrated NbN-SSPD on the same chip. By measuring the autocorrelation function of similar QDs using off-chip detection and simultaneously showing on-chip detected resonant fluorescence, we prove the non-classical character of the detected light. The results of this prototype quantum optical circuit provide a test bed for all-optical generation, distribution and detection of non-classical light on a common photonic chip.

**Methods**

**Sample preparation:** The samples investigated were grown using solid source molecular beam epitaxy and consisted of a $350\,\mu m$ thick GaAs wafer onto which a $2\,\mu m$ thick $Al_{0.8}Ga_{0.2}As$ waveguide cladding layer was deposited. Following this, a $250\,nm$ thick GaAs waveguide core was grown into which a layer of self-assembled InGaAs QDs was embedded at its midpoint. The growth conditions used resulted in dots with a typical lateral (vertical) size of $25 \pm 5\,nm$ ($5 \pm 1\,nm$) as shown in the top-right inset of fig 1(a), an areal density of $6 \pm 1\,\mu m^{-2}$ and PL emission around $\sim 1348\,meV$ at $4\,K$ with an inhomogenously broadened ensemble linewidth of $83\,meV$. After growth, the native oxide was removed from the sample surface using an HCl dip and a high quality $10 \pm 0.5\,nm$ thick NbN superconducting film was deposited using DC reactive magnetron sputtering. By carefully optimizing the deposition temperature, rate and the Nb:N ratio, high quality superconducting films were obtained on the GaAs substrate ($T_C = 11.9 \pm 0.2\,K$), despite the $26\%$ lattice mismatch[41,42]. To suppress the scattering of laser stray light into the detector, the backside of the sample was chemically and mechanically polished using a 5% $Br_2$:$CH_3OH$ solution and an absorbing Si layer was grown on the backside of the sample using electron beam evaporation. As described in detail in the supplementary material section S6, scattered laser light on the SSPD is only an issue for excitation energies below the GaAs bandgap for which the substrate is transparent. For such below gap excitation scenarios, the signal-to-background ratio ($\eta$) reduces significantly from $\eta \geq 420$ for above gap excitation, to $\eta \leq 3$ for sub-bandgap excitation. To circumvent this effect, the polishing of the sample backside was employed to reduce the RMS roughness to $< 20\,nm$ decreasing the diffuse light scattering by one order of magnitude. Moreover, we coated the backside of the wafer with a $1\,\mu m$ thick amorphous Si-layer to reduce the index contrast at the rear surface and efficiently absorb photons with $E_{exc} \geq 1.2\,eV$ at the back side of the sample (for details see supplementary material section S6). Whilst these combined measures increased $\eta$ by more than two orders of magnitude, the obtained values remained too high for on-chip detection of QD luminescence via the SSPD. To further suppress the direct detector illumination by laser stray light originating from the top, a sequence consisting of $50 \pm 15\,\mu m$ Teflon, $12 \pm 0.5\,\mu m$ Aluminium foil, $50\,\mu m$ Teflon foil was carefully placed over the detector under an optical microscope. The complete sandwich structure (shown in the supplementary material section S5) was then fixed to the side of the chip carrier.

The nanowire detectors were then defined using electron beam lithography with a negative tone resist and reactive ion etching using a $SF_6$ / $C_4F_8$ plasma to form an NbN nanowire meander consisting of $18x$, $100 \pm 5\,nm$ wide nanowires separated by $150 \pm 5\,nm$ to form a detector with a width of $4.5\,\mu m$ and a total length of $30\,\mu m$ along the waveguide axis. A scanning electron microscope image of the resulting NbN nanowires on GaAs is presented in fig 1(a) – inset at the bottom. Subsequently $\sim 2600\,\mu m$ long, $20\,\mu m$ wide multimodal ridge waveguides were defined using photolithography and wet etching in a citric acid + $H_2O_2$ solution. The waveguides were defined such that a pair of nanowire detectors is centred on one end, penetrating $\sim 60\,\mu m$ into the waveguide end to ensure optimum evanescent coupling[14].

**Device operation and characterization:** The SSPD was operated at liquid helium temperatures inside a cryogenic dip stick. A bias-tee was used to drive a fixed bias current of $I = 0.95 \times I_c = 15.0$ μA through the nanowires. A low noise voltage source in series with a $50\,k\Omega$ resistor was operated as a constant current source. Voltage pulses arising from single photon detection events were then amplified using two $30\,dB$ high-bandwidth amplifiers and detected with a PicoQuant Timeharp Card 200 time correlated counting module recording histograms of the time intervals between the trigger signal provided by a microwave source and the voltage pulse arising upon photon detection. A continuous wave diode laser source was sent through a Mach – Zehnder electro-optical modulator driven by a $3.3\,GHz$ microwave source. In this way, the laser output was temporally gated into $290 \pm 70\,ps$ long Gaussian pulses, as shown in fig 1(c), at a repetition rate of $100\,MHz$ and a duty-cycle of $3\%$. Using this technique, we temporally filtered the on-chip detected SSPD signal enabling us to separate the scattered laser light from the incoherent part of the QD luminescence signal detected by the SSPD. A detailed description of the complete experimental setup is given in the supplementary material (section S1). For all off-chip detected confocal PLE measurements the laser was guided through a pulse shaper with a bandwidth of $1\,nm$ suppressing laser sidebands. The light emitted by the QDs was collected with a single mode optical fiber and guided to a $0.5\,m$ focal length spectrometer. Additionally, a $975\,nm$ short/long – pass filter was used in the excitation/detection path. In addition, for the broad PLE sweep presented in fig 2(b), a $980\,nm$ band-pass filter was employed.

## Acknowledgements

We gratefully acknowledge D. Sahin, A. Fiore (TU Eindhoven) and K. Berggren, F. Najafi (MIT) and R. Hadfield (University of Glasgow) for useful discussions and the BMBF for financial support via QuaHL-Rep, project number 01BQ1036, and Q.com via project number 16BQ1036, the EU via the integrated project SOLID and the DFG via SFB 631-B3 and the ARO (grant W911NF-13-1-0309).

## Author contributions

G.R., F.F., M.K. and J.J.F. conceived and designed the experiments. G.R. prepared the samples. G.R. and F.F. performed the experiments and analyzed the data. S.L. assisted with the autocorrelation measurement. A.A., K.M. and R.G. contributed materials. G.R., F.F., M.K. and J.J.F. prepared the manuscript. All authors reviewed the manuscript.

## Additional information

7 pages of supplementary information is available showing a description of the experimental setup, detailed PL studies of $X_1$ and $X_3$, an analysis of the gating laser power dependence, the SSPD operation parameters, the sample structure and measurements concerning the stray light suppression.

## Competing financial interests

The authors declare no competing financial interests.

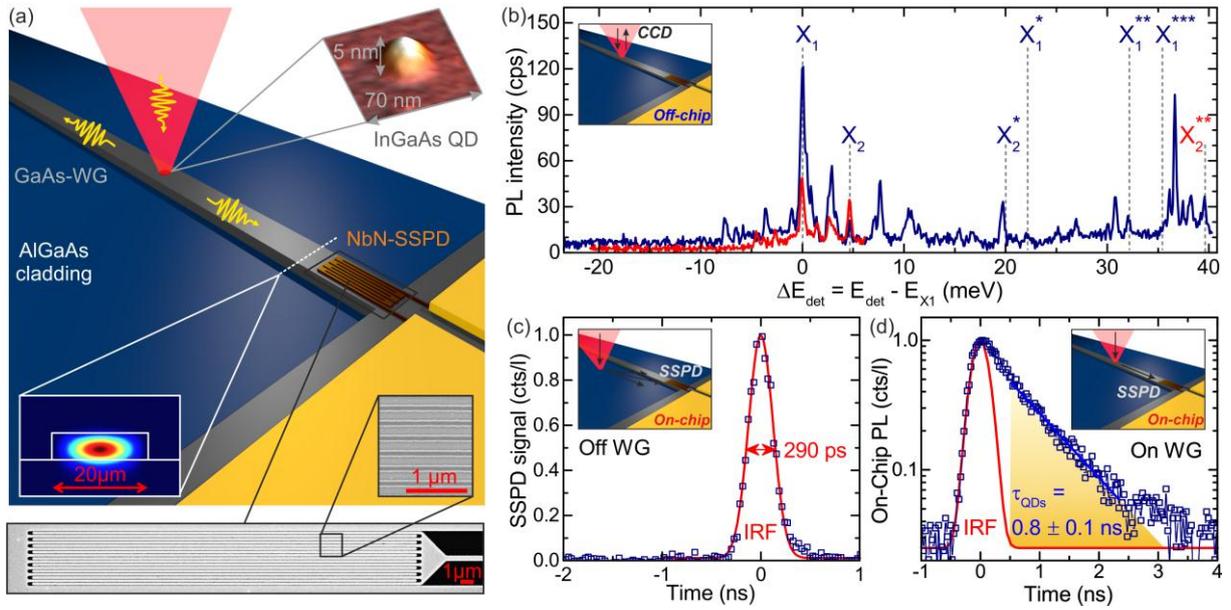

**Figure 1.** (a) Sample structure combining a $2\,mm$ long GaAs ridge waveguide (FDTD simulation of the fundamental mode profile as an inset) containing a single layer of InGaAs QDs, as shown in the AFM image on the top right, that are evanescently coupled to an NbN-SSPD. An SEM image of the SSPD is shown on the lower panel. (b) PL intensity detected in confocal geometry for excitation in the wetting layer at $1347.7\,meV$ (blue curve) and at $X_2^{**}$ (red curve). In the text we focus on the two quantum dot transitions labelled $X_1$ and $X_2$. The corresponding excited state transitions that we focus on are labelled $X_{1,2}^*$, $X_{1,2}^{**}$ and $X_1^{***}$ respectively. (c) Instrument response function (IRF) of the electrically modulated excitation laser, detected on-chip using the SSPD. The signal is plotted in units of counts per detected laser photon (cts/l). (d) Typical time transient of multiple quantum dots excited in the wetting layer and detected on-chip using the integrated SSPD. The IRF is shown in red, whereas the integration window used to extract the intensity originating solely from the QD emission is indicated by the yellow shaded region.

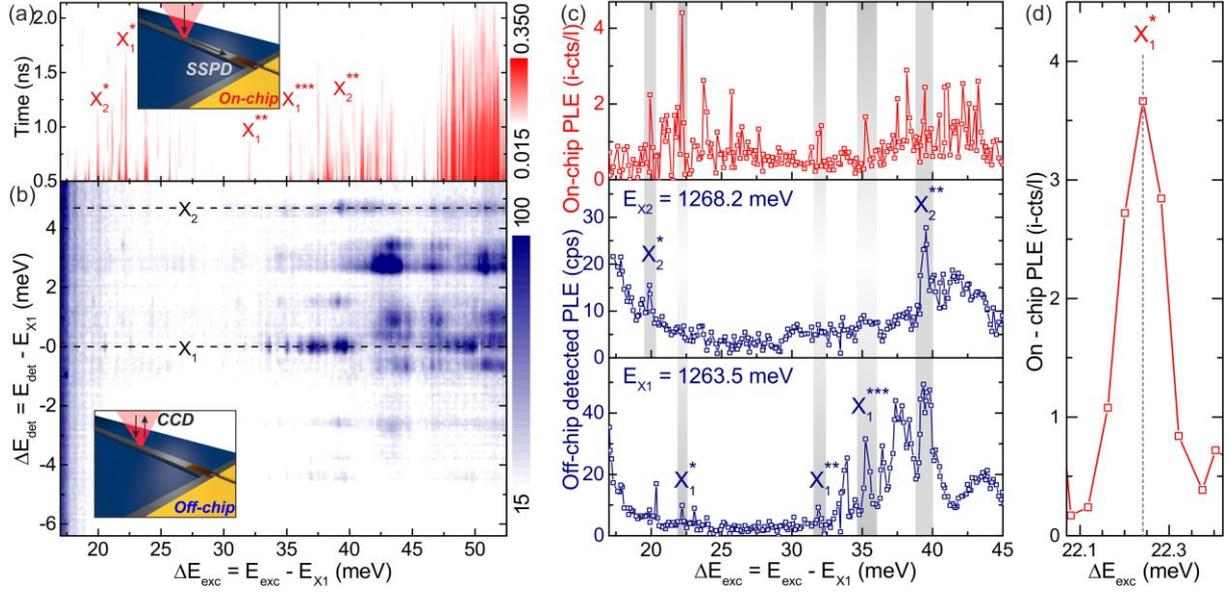

**Figure 2.** (a) Time-resolved on-chip PL intensity as a function of energy difference $\Delta E_{exc}$ between the laser $E_{exc}$ and the $X_1$ transition $E_{X1}$. The lines that we focus on are marked $X_{1,2}^*$, $X_{1,2}^{**}$ and $X_1^{***}$. (b) Spectrally filtered, off-chip detected PL intensity for the same range of excitation energy as probed in the on-chip measurement. The two highlighted lines $X_1$ and $X_2$ correspond to the QD transitions presented in fig 1b. (c) Top panel: Integrated on-chip PL intensity for times > 0.5 ns plotted as a function of $\Delta E_{exc}$. The on-chip signal is plotted in units of integrated counts per detected laser photon (i-cts/l). Middle/bottom panel: PL intensity detected in off-chip geometry along the two lines marked in fig 2b. The 5 common resonances that we focus on are highlighted by the grey shaded regions and labeled with respect to the spectrum shown in fig 1b. (d) High-resolution scan with on-chip detection of resonance $X_1^*$.

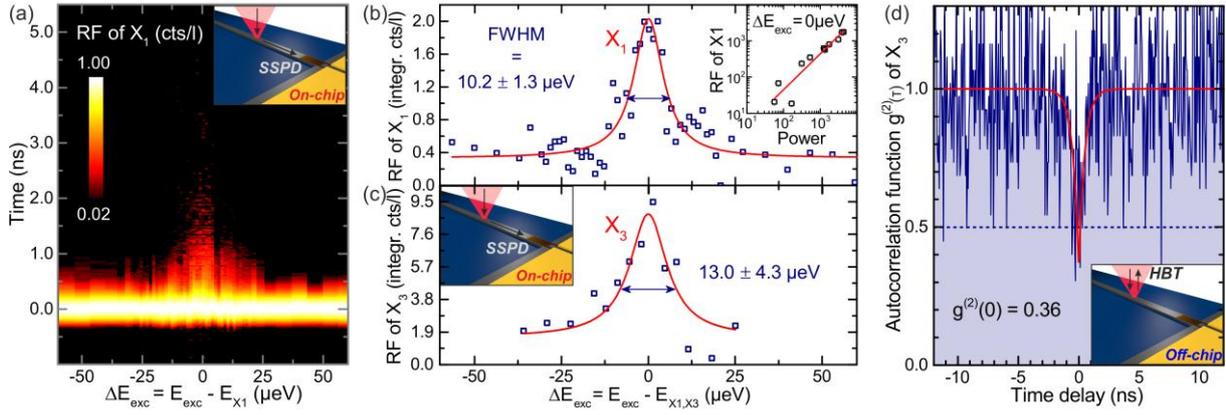

**Figure 3.** (a) Time-resolved resonant fluorescence signal obtained from $X_1$, as detected in on-chip geometry using the integrated SSPD. The data is color coded in counts per detected laser photon (cts/l). (b) and (c) Integrated RF signal of $X_1$ and $X_3$ for a time-window from $0.5$ to $5.0$ ns (blue squares) showing a Lorentzian line-shape (red solid line) with a FWHM of $10.2 \pm 1.3$ μeV and $13.0 \pm 4.3$ μeV, respectively. Inset in (b): Linear power dependence of the RF signal for $0$ μeV laser detuning. (d) Autocorrelation function $g^{(2)}(\tau)$ measured for QD transition $X_3$ with an emission energy of $E(X_3) = 1296$ meV under quasi-resonant continuous wave excitation at 1362 meV shown in blue. The deconvoluted fit to the data is shown in red, resulting in $g^{(2)}(0) = 0.36$.

1.

# On-chip generation, routing and detection of quantum light

## SUPPLEMENTARY INFORMATION

G. Reithmaier[1], M. Kaniber*[1], F. Flassig[1], S. Lichtmannecker[1], K. Müller[1,3], A. Andrejew[1], J. Vučković [3], R. Gross[2,4] and J. J. Finley*[1,4]

[1]Walter Schottky Institut und Physik Department, Technische Universität München, Am Coulombwall 4, 85748 Garching, Germany.

[2]Walther-Meißner-Institut, Bayerische Akademie der Wissenschaften und Physik-Department, Technische Universität München, 85748 Garching, Germany.

[3]E. L. Ginzton Laboratory, Stanford University, Stanford, CA 94305, USA.

[4]Nanosystems Initiative Munich (NIM), Schellingstraße 4, 80799 München, Germany.

*Correspondence to finley@wsi.tum.de or kaniber@wsi.tum.de

Content:

# S1 Experimental setup

Figure S1 shows a schematic overview of the components used in the on- and off-chip detected electro-optical experiments described throughout the manuscript.

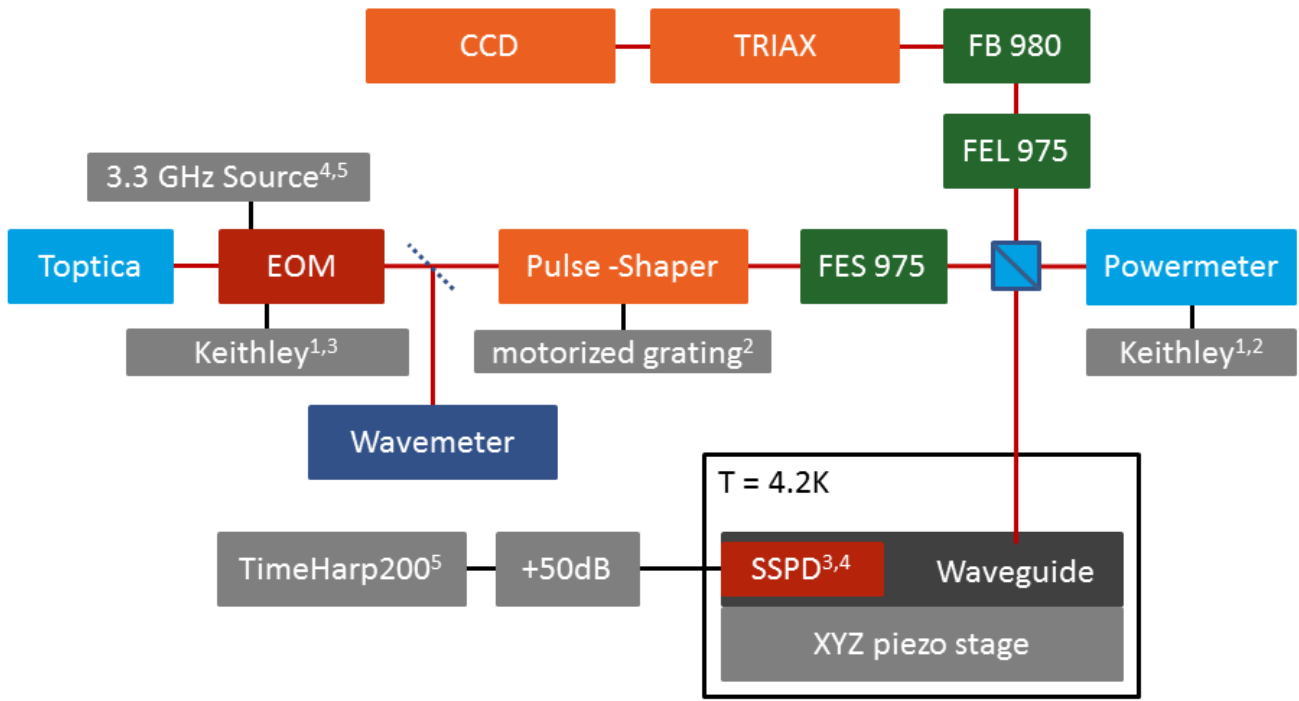

**Figure S1:** Optical components to prepare laser pulses. Toptica: cw-tuneable diode laser, EOM: electro-optical modulator (Mach-Zehnder interferometer), Wavemeter: Wavelength meter with 600MHz accuracy, Pulse-shaper: Pulse-shaper in near-Littrow configuration with 1200/mm blaze grating, FES 975 (FEL975): 975nm OD4 shortpass (longpass) filter, FB980: 980nm OD4 bandpass filter, TRIAX: Spectrometer, SSPD: On-chip superconducting single photon detector. Superscript numbers identify communicating devices

A cw-tuneable diode laser (Toptica) is sent through a Mach-Zehnder interferometer (EOM) which is operated at a bias voltage controlled via a voltage source (Keithley) and modulated via a 3.3 $GHz$ pulse generator. A wavelength meter (600 $MHz$ accuracy) is attached via a pellicle to log the exact wavelength. A pulse-shaper in near-Littrow configuration and a 975 $nm$ OD4 shortpass filter (FES 975) is used to block laser sidebands. The filtered laser pulse is coupled into a single mode fiber and sent into the excitation path of the dipstick cryostat. The laser beam power is read out via a 50:50 unpolarised beamsplitter and a powermeter. The other half of the laser beam is sent into the cryostat and focused on the sample containing the waveguide and on-chip superconducting detector (SSPD) being operated at 4.2 $K$. Reflected light is filtered by a 975 $nm$ OD4 longpass filter (FEL975) and a 980 $nm$ OD4 bandpass filter (FB 980) and then analysed with an imaging monochromator (TRIAX + CCD). Light coupled into the waveguide is detected by the on-chip SSPD, enhacned via two amplifiers ($+ 50\ dB$) and read-out with a time correlated single photon counter (TimeHarp200).

Superscript numbers indicate equipment communicating with each other. The GHz pulse generator is used to allow for time-resolved measurements on the SSPD and is giving the trigger signal for the TimeHarp. To find the optimum bias voltage of the EOM, the laser stray light in the SSPD is analysed and automatically optimised with respect to the pulse to cw ratio of the laser. In a pure cw confocal measurement, the bias point of the EOM is automatically optimized to maximize power transmission of the EOM by reading out the power on the powermeter. When performing a wavelength scan, the motorized grating of the pulse shaper can both be used to maximise transmitted laser power or to set a constant power level.

## S2 Power dependent PL characterisation

In the following section, PL spectra of the emission lines $X_1$, $X_2$ and $X_3$ are presented as well as an analysis of the power dependency of these emission lines. The experiments were performed in confocal off-chip detection under wetting layer excitation at $E_{exc} = 1347.7\ nm$.

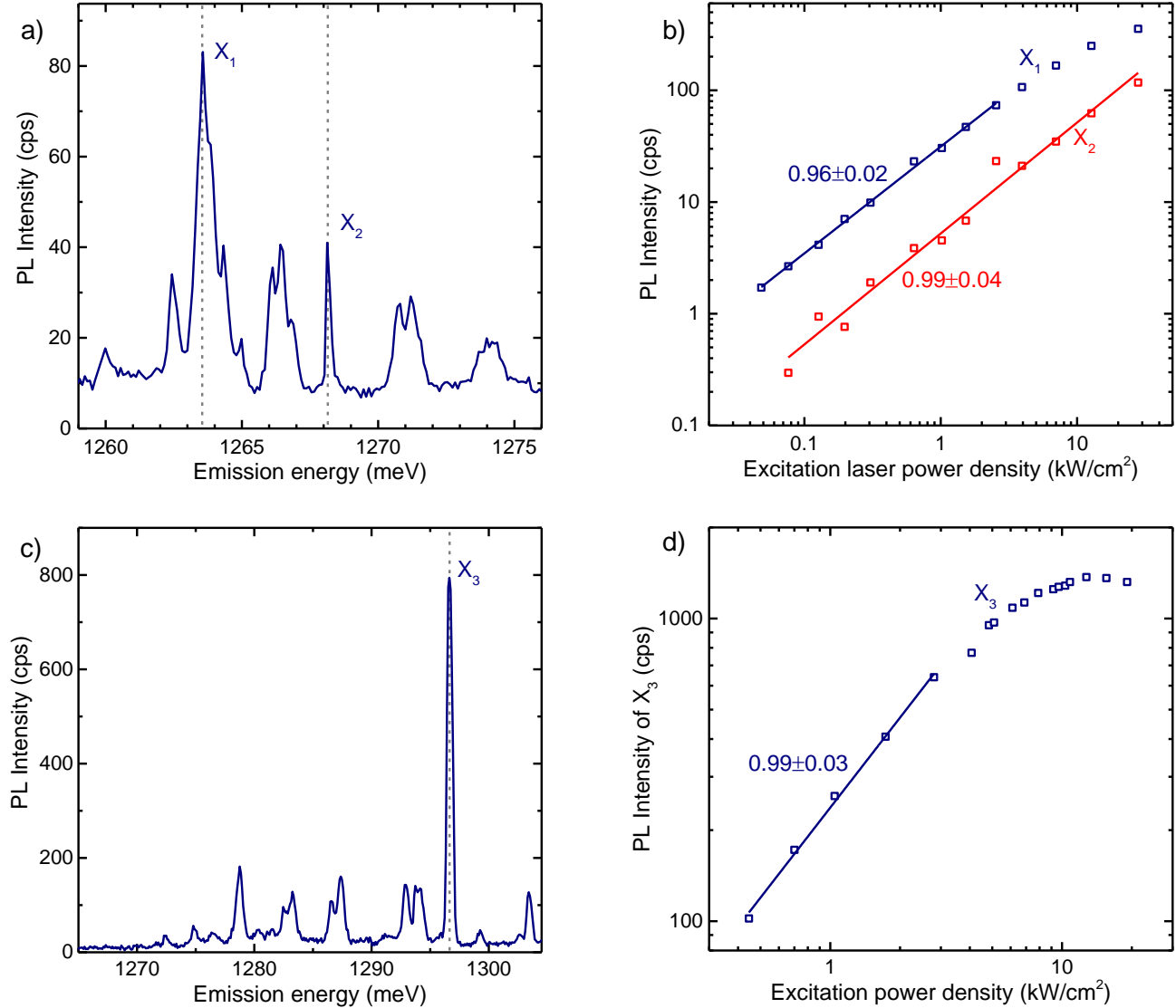

**Figure S2:** (a) PL signal for an excitation energy of $1347.7\ meV$ and an excitation power density of $2.55\ kW/cm^2$. The features marked $X_1$ and $X_2$ refer to the QD emission lines discussed in the main text. (b) PL intensity as a function of excitation power density for $X_1$ and $X_2$ in double-logarithmic plot. (c) PL signal of $X_3$ for an excitation energy of $1347.7\ meV$ and an excitation power density of $4.08\ kW/cm^2$. The feature marked as $X_3$ refers to the QD emission line discussed in figure 3(c) in the main text. (b) PL intensity as a function of the excitation power density for $X_3$ in double-logarithmic plot.

Figure S2(a) shows the off-chip detected PL intensity of emission lines $X_1$ and $X_2$, whereas figure S2(c) shows a spectrum of $X_3$. The data was recorded for an excitation power density of $2.55\ kW/cm^2$ for $X_1$ / $X_2$ and $4.08\ kW/cm^2$ for $X_3$, respectively. Our studies of the power dependency of the QD emission is presented in a double-logarithmic scale in figures S2(b) and (d) for $X_1$ / $X_2$ and for $X_3$, respectively. Here, all observed features show linear power dependences with the exponent of the linear fits being very close to 1.0 as depicted in the plots, indicating that those features originate from single excitonic transitions.

## S3 Gating laser analysis

In order to observe resonant fluorescence in samples without a built-in potential, a weak non-resonant laser is mixed into the excitation to stabilize the electrostatic environment[31,35]. The effect of the above-bandgap gating laser on the detected RF signal is presented in this section.

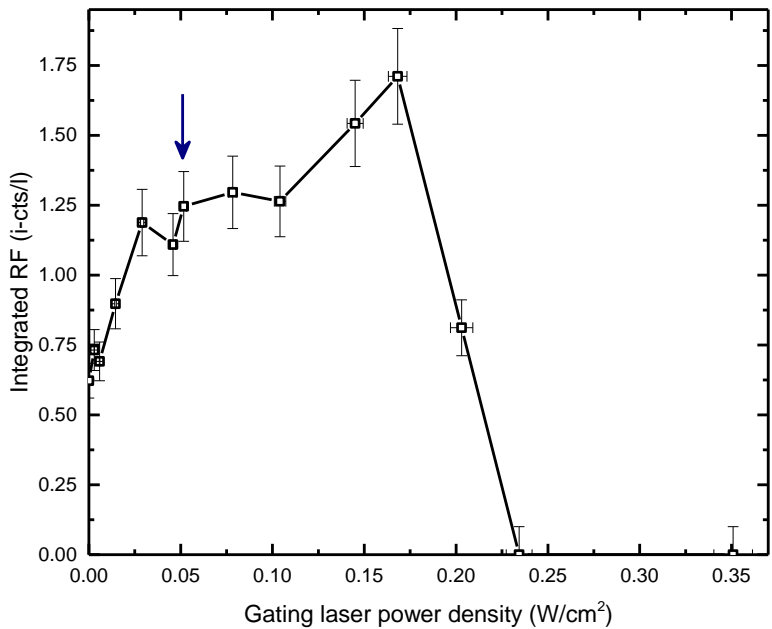

**Figure S3:** Integrated RF intensity as measured under resonant pulsed excitation for different gating laser powers. The resonant laser power was kept constant while the defocused gating laser power was varied. The blue arrow marks the power density ($0.06 \, W/cm^2$) used for RF scans. The signal is given in integrated counts per detected stray laser photon.

Figure S3 shows the integrated RF peak amplitude for the $X_1$ feature as a function of the gating laser power. The power of the excitation laser being in resonance with the $X_1$ transition was kept constant. Without applying the gating laser, only very little ($0.62 \, i-cts/l$) PL signal was observed which was strongly increased by mixing in an above-bandgap mixing laser with an energy of $1534.5 \, meV$. However, this gating laser caused additional, non-resonantly excited PL background adding up to the cw-part of the diode laser stray light. Both, the PL background and the cw-part contribute to background signal at the detector, thus with higher gating laser powers the signal to background ratio decreases. Therefore, a gating laser power slightly below the maximum integrated RF was chosen ($0.06 \, W/cm^2$). At even higher powers, the RF signal drastically reduced until it completely vanished. The steep decrease can be explained by the fact that the non-resonantly created PL background signal gets orders of magnitude larger than the detected RF signal causing the temporal filtering routine to fail. Furthermore, with increasing non-resonant excitation, the QD $X_1$ transition gets populated, thus reducing the RF signal observed from this feature.

The different dots that were investigated showed a similar behaviour, yet with varying gating laser powers required for optimum signal. Thus, an ideal operating point has to be found for each individual QD.

The gating laser was strongly defocused with a spot size of 21 µm, as it was mixed via the confocal detection path in the on-chip experiments. This path was focussed for light of $1305 \, meV$ and, thus, the gating laser with an energy of $1534.5 \, meV$ was defocussed. By mixing the laser through a second optical path with an individually tuneable focal plane, RF experiments with a focused gating laser spot were done, however, resulting in a lower RF signal. Therefore, we conclude that a large size illumination with a weak non-resonant laser $\sim 0.06 \, W/cm^2$ gives the best stabilizing effect and still does not create too much background light for on-chip experiments.

# S4 SSPD characteristics

A stable operation of the superconducting single photon detector (SSPD) is crucial for all on-chip measurements discussed in the main text. For a SSPD this means providing a stable temperature, a stable and noise-free electrostatic environment and a well-chosen bias current. Therefore, the device is operated in a dipstick cryostat, whilst being operated at 0.95 $I_C$, where $I_C$ is the device critical current.

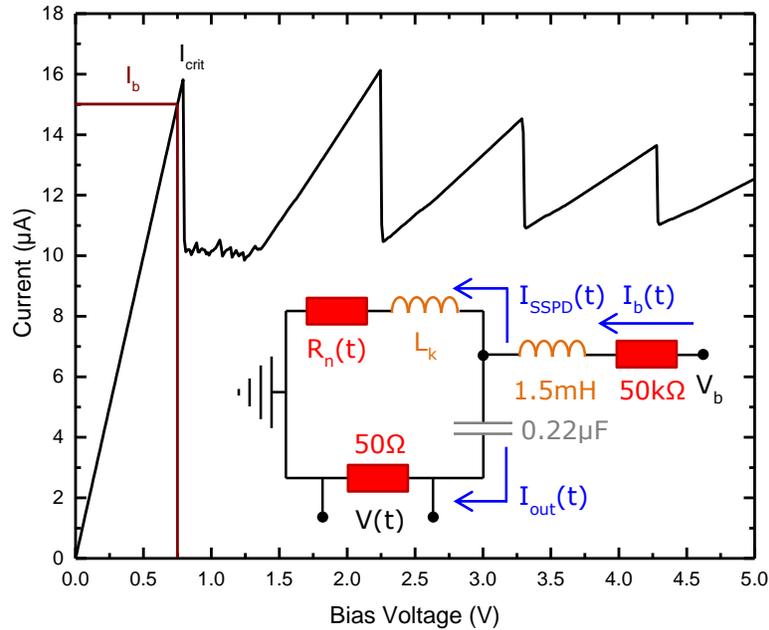

**Figure S4:** Voltage current sweep of the SSPD. $I_c$: critical current of the detector, $I_b$, dark red: bias current at which the detector is operated. Inset: electric circuit for operating the SSPD

A voltage-current sweep of the integrated SSPD at $4.2\ K$ is presented in figure S4. The linear increase to $0.75\ V$ corresponds to the regime where the detector is in the superconducting state and the resistivity is dominated by the $50\ k\Omega$ bias resistor. At $I_C = 15.8 \pm 0.1\ \mu A$ the current density at the dominant constriction exceeds the critical current density, causing a breakdown of the superconducting state. Therefore, an additional resistivity and instant decrease of the bias current is observed at the point marked $I_C$. The point of operation of the detector is chosen to be at $I_b = 0.95\ I_C = 15\ \mu A$ (dark red line in figure S4). This is a trade-off between detection efficiency, dark counts and operational stability[18]. Insert: Electric circuit for operating the detector. A bias voltage $V_b$ is applied to the $50\ k\Omega$ resistor giving rise to a current $I_b$ which is then sent through a bias tee consisting of an inductance and a capacitor. The SSPD is connected at the AC+DC port of the Bias-Tee (the detector is represented by an inductivity $L_K$ and a time-dependent resistance $R_N(t)$). In case of an event in the SSPD, a time-dependent voltage transient $V(t)$ is created in the AC output arm. This signal is then amplified (not drawn in the schematic) and then guided to the read-out electronics.

# S5 Sample structure

Figure S5 (a) shows a schematic of the sample layer structure. On top of the 350 µm thick GaAs wafer (medium grey) a 2 µm thick layer of $Al_{0.8}Ga_{0.2}As$ (blue) was grown that acts as a waveguide cladding for the 250 nm thick GaAs ridge waveguide (medium grey) on top. The layer of self-assembled QD's is marked by red dots in the waveguide centre, whereas the colour map indicates the simulated intensity distribution of the fundamental transverse electro-magnetic mode guided in the waveguide. The NbN nanowire detector is shown in orange. A 1 µm thick layer of amorphous Si (dark grey) is grown on the back side of the mechanically polished sample. This layer is included to absorb stray laser light at the back side of the sample that otherwise would get reflected and diffusely scattered at the sample backside. The sample top is covered by a sandwich structure of Teflon (PTFE, light blue) and aluminum (light grey). The latter one shields off any direct photons incident on the detector, while the former one acts as a pair of insulating layers to prevent short-circuits caused by the aluminum.

Figure S5 (b) schematically shows the corresponding refractive indices for photon energies $E_{ph} \sim 1.3\ meV$. Si has a small imaginary part of the refractive index which is not visible on this scale. As shown in figure S6(a), the amorphous Si layer works as an absorber, however we note that the use of a sputtered layer of NbN might show even better absorption.

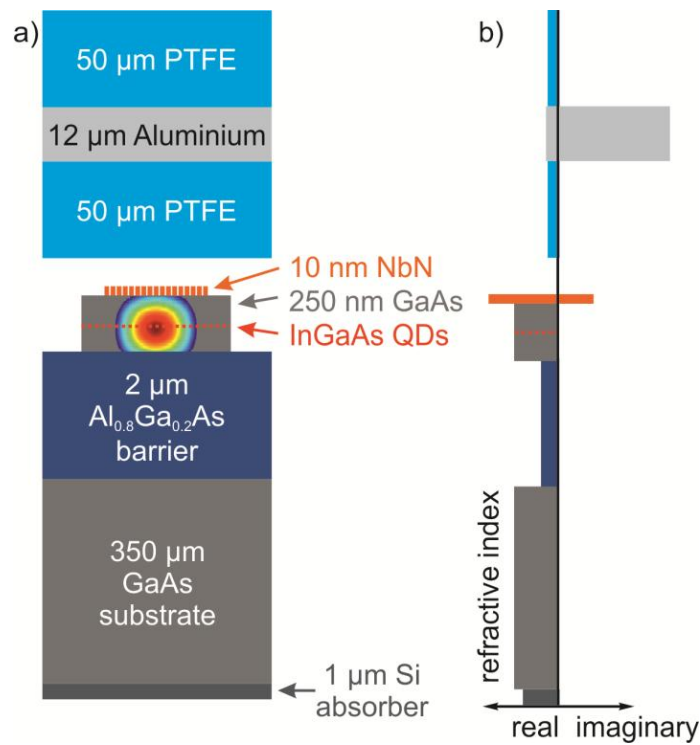

**Figure S5:** (a) Schematic layer structure of sample (not to scale). Fundamental mode profile simulation in the 250 nm thick ridge waveguide shown as a color map. (b) Real and imaginary parts of refractive indices of the sample structure.

# S6 Stray light suppression

When testing quasi-resonant and above-bandgap excitation[28], the signal on the detector was dominated by laser stray light. Its origin could be assigned to laser light reflected at the back side of the sample and guided into the detector. Hence, the back side was polished and coated with Si and longer waveguides were introduced. These optimizations showed a decrease of stray light by a factor of 500, when measuring the stray light in a simple room temperature, angle-dependent reflectivity setup. The polishing process was a chemical-mechanical polishing using a $5\%\ Br_2:CH_3OH$ solution to reduce the sample backside surface roughness to $< 20\ nm$. Afterwards, the backside was coated with $\sim 1\ \mu m$ of Si using electron beam evaporation deposition.

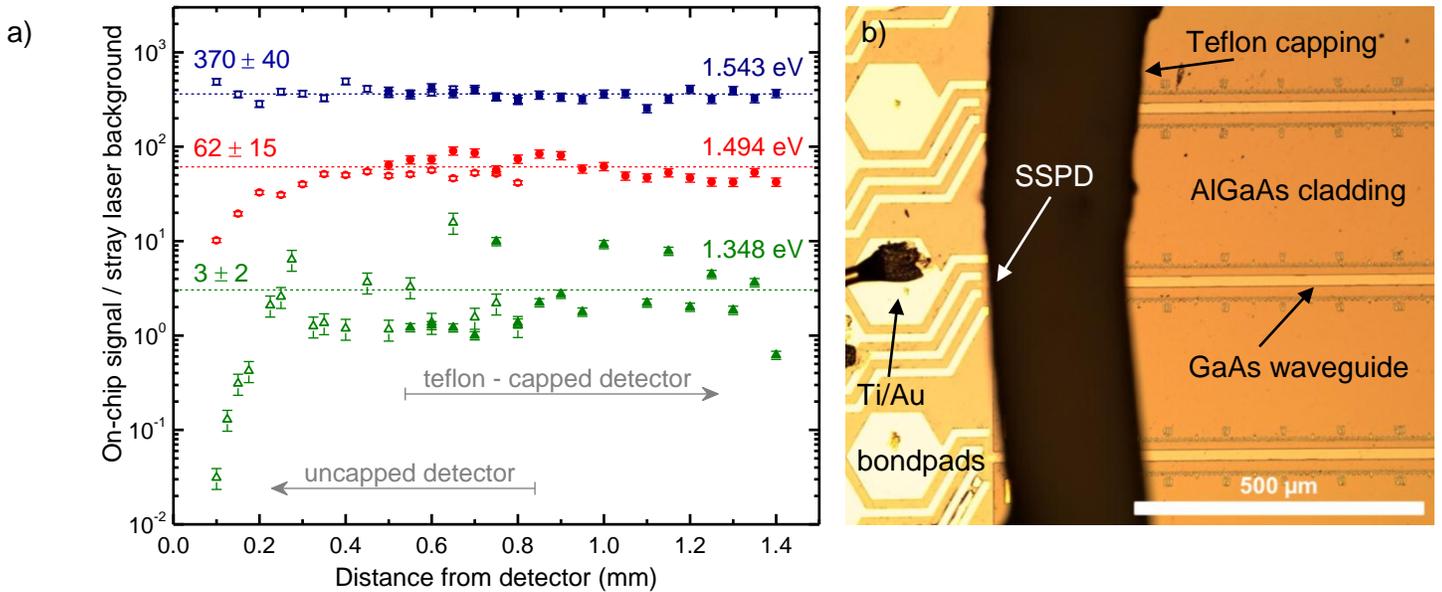

**Figure S6:** (a) Signal to stray light ratio for different illumination energies for both uncovered and covered detectors for illumination at a power density of 1273W/cm² (b) Image of sample with a PTFE-Aluminum-PTFE stripe (black) that covers the detectors. Contact pads (gold) are shown on the left with a bonding wire and waveguides with distance markers are depicted on the right.

Figure S6 (a) shows the excitation-detector distance dependency of the on-chip PL signal to stray light ratio for uncapped (open symbols) and Teflon – capped (full symbols) detector illumination scenarios for different energies. The on-chip signal is given by the SSPD count rate when illuminating the waveguide. Here, both QD luminescence and laser stray light is collected. The stray light is independently measured by the count rate when illuminating the AlGaAs barrier besides the waveguide, as shown in the microscope image in figure S6(b).

The strong increase of the signal to stray light ratio for illumination within the quasi-continuum QD states at $1.348\ eV$ (green data points) when increasing the distance to $300\ \mu m$ is caused by the absorption of stray light at the sample backside. However, the signal to stray light ratio stays then constant at a very low value of $3 \pm 2$ for larger distances. This qualitative behaviour is also observed for an excitation energy of $1.494\ eV$ (red data points) but with a larger overall signal to background ratio of $62 \pm 15$ stemming from the higher effectivity of charge carrier generation, as the excitation energy is within the wetting layer states. When considering illumination above the bandgap at an energy of $1.543\ eV$ (blue data points) the signal to background ratio stays constant at a value of $370 \pm 40$ over the whole distance range investigated. The fact that the signal to background ratio is much higher when compared to the below band gap excitation scenarios as well as the complete absence of an increase for distances $< 300\ \mu m$ stems from the highly efficient laser stray light absorption in the substrate for this above band gap illumination.

As the signal to background ratio of $3 \pm 2$ for an excitation within the excited state transitions of the QDs is not sufficient to do excitation energy dependent on-chip PL spectroscopy, we further distinguish between laser light and QD PL with a temporal filtering technique, as described in the main text related to figure 1(c) and (d).